\documentclass[aps,floatfix]{revtex4}
\usepackage{latexsym}
\usepackage{graphicx}

\begin{document}

\title{Five-dimensional warped product space-time with time-dependent warping and a scalar field in the bulk}

\author{Sarbari Guha}
\affiliation{Department of Physics, St. Xavier's College (Autonomous), Kolkata 700016, India}
\author{Pinaki Bhattacharya}
\affiliation{Gopal Nagar High School, Singur 712409, West Bengal, India\\
Department of Physics, Jadavpur University, Kolkata 700032, India}
\begin{abstract}
We have considered gravity in a five-dimensional warped product space-time, with a time-dependent warp factor and a
time-dependent extra dimension. The five-dimensional field equations are derived for a spatially flat FRW brane and the energy conditions and the nature of bulk geometry have been examined. It is found that the expansion of the four-dimensional universe depends on its location along the extra dimension and will be different at different locations in the bulk spacetime. At low energies, the trapping of fields within the brane implies a specific correlation between the warp factor and the extra-dimensional scale factor. Generally, the bulk is not conformally flat. At high energies, the bulk is assumed to be sourced by a scalar field with self-interaction. The analysis shows that the potential of the scalar field source of gravity at a given position along the fifth dimension, is related to the Hubble parameter on the brane at that position in the bulk.
\end{abstract}

\maketitle

\section{Introduction}
In recent theories of particle physics \cite{AKA1,Rubakov,GSW,Pol} the universe is regarded as a
(3+1)-dimensional spacetime embedded in a higher-dimensional bulk. The knowledge that the effective five-dimensional
(5D) theory of the strongly coupled heterotic strings is a gauged version of the ($N=1$) 5D
supergravity with four-dimensional (4D) boundaries \cite{Lukas1,Lukas2}, led to extensive studies in this field. Inspired by the D-branes of string theory, Arkani-Hamed, Dimopoulos and Dvali \cite{ADD1,ADD2,ADD3,ADD4} and Randall and Sundrum
(RS) \cite{rs} proposed theories of extra dimensions in which some of these fields (which may sometimes
represent only gravity) propagate in the extra dimensions, while standard model (SM) fields remain confined
to a ($3+1$)-dimensional submanifold, called {\it brane}. The simplest illustration is available in the framework of a 5D theory with a
single warped extra dimension \cite{lrr,MS,Mannheim}. The issue of localization of gravity and particles was
addressed by many researchers \cite{rs,loclzn1,loclzn2,loclzn3,loclzn4,loclzn5,loclzn6,loclzn7,loclzn8,loclzn9,loclzn10,loclzn11,RPU}. Cosmological consequences have been studied in the RS model of both
the single brane and two brane versions \cite{bdl1,bdl2,others1a,others1b,others2a,others2b,others3a,others3b,others4a,others4b,others4c}.

The extra-dimensions play a crucial role in a dynamical space–time \cite{AKA2}. The bulk is ``slightly warped'' in the
models of heterotic M theory with the inter-brane distances being represented by scalar fields (called \emph{moduli}),
introduced through the integration constants in the classical solutions. For solutions over a continuous range
of integration constants, the moduli are massless, giving rise to long range scalar forces, which evolve cosmologically,
being time-dependent. This leads to time-variations of the fundamental constants \cite{varmod1,varmod2,varmod3},
indicating that at least some moduli fields are not stabilized. To avoid this, the moduli fields are assumed to be stabilized
by some mechanism in the early universe \cite{early1,early2,early3}. The RS model is ``highly warped'' and devoid of
any bulk scalar. But string theory requires the dimensionally reduced low energy effective action to include both
the higher-order curvature terms, as well as dilatonic scalar fields \cite{Lukas1}. Thus scalar fields were
introduced in the bulk of the RS model to stabilize the extra dimension \cite{GW}, with the single modulus,
called the ``radion'', being related to the thickness of the AdS slice.

In the RS model, the bulk and the branes are maximally symmetric. The fields living on the negative tension brane, have
masses proportional to the radion expectation value and for generic matter in the bulk, the one loop effective potential
is not sufficient to stabilize the radion at a phenomenologically acceptable mass \cite{GPT}. In the warped braneworld
models which are not maximally symmetric, the bulk contains a dilaton field, with an
exponential potential coupled to the 5D gravity. The solutions have a power-law warp factor which yields a
power-law inflation on the brane \cite{powerlaw,KT1}, although the bulk itself is not inflating. The back reaction of the dilaton coupling in the bulk to the 4D brane gives rise to the non-vanishing component of the energy-momentum tensor along the fifth dimension \cite{KKOP}.

In the RS1 setup, in presence of a bulk scalar field which couples to the branes, the inter-brane distance is not arbitrary
and the expected massless mode acquires a nonzero mass, so that Einstein gravity is recovered \cite{scalar1,scalar2,scalar3,scalar4,scalar5,scalar6,MK}. In the RS2
setup, although, the hierarchy problem cannot be solved, there are several important consequences
\cite{DGP1,DGP2,Deffayet,rs2cosmo1a,rs2cosmo1b,rs2cosmo1c,rs2cosmo2a,rs2cosmo2b,rs2cosmo2c,rs2cosmo2d,rs2cosmo2e}. The bulk scalar facilitates to generate the hypersurface in the form of the so-called \emph{thick brane} \cite{BCG}.
Although SM fields cannot move freely through the extra dimensions, they may be able to access the extra dimensions to
distances $\sim m^{-1}_{EW} \sim 10^{-19}$ m, without contradicting the Standard Model \cite{ADD1,ADD2,ADD3,ADD4}. In such a case, the braneworld should be a field-theoretic construction, in the form of a topological defect (a domain wall (DW)), in a higher dimensional space, an idea originally
proposed by Akama \cite{AKA1} and Rubakov and Shaposhnikov \cite{Rubakov}. A simple way to construct a thick brane
is to replace the delta functions in the action by non--singular source functions. Generally, a scalar field supported by a potential is
responsible for the generation of a DW, providing smoothed out versions of the original
delta-function defects in the warped RS model \cite{Gremm,Giovannini2,Bronnikov1,Bronnikov2,DFGK1,DFGK2}. Alternatively one can also use self-interacting scalar fields minimally coupled to gravity. A more elaborate method is to introduce a non-minimal coupling between gravity and matter \cite{ahmedgrad1,ahmedgrad2,ahmedgrad3,Farakos1,Farakos2}.

Many authors have constructed generalized solutions to build realistic models
for addressing the unsolved issues in astrophysics and cosmology \cite{SMS1,SMS2,Greg1,Greg2,Greg3,nogo1,nogo2,GC1,GK1a,GK1b,Gog}. In the
dynamical brane solutions in supergravity and string theory \cite{sugra1,sugra2,sugra3,sugra4,sugra5,sugra6,sugra7}, the warp factors can be time-dependent, depending on the ansatz
for the fields. After compactification of the internal space, the solutions lead to Friedmann-Lema$\hat{i}$tre-Robertson-Walker (FLRW) cosmology with power-law expansion. Extension of \cite{ahmedgrad1} to spherically symmetric time-dependent solutions was done for a 5D canonically coupled scalar field, and the cosmology of such models was studied \cite{Bernardini}. The equivalence between Born–Infeld and effective real scalar field theories in warped geometry, and localization of matter have also been investigated \cite{BB,BR}.

The concept of time-dependent soliton solutions arising typically in the (1+1)-dimensional spacetimes was extended to 5D
by Giovannini \cite{Giovannini}. Such solutions are appropriate for modelling the 5D domain walls, with infinite extra dimension, and can be used to describe higher-dimensional cosmological models, for a fixed value of the bulk radius, which arise from thick branes. The so-called gravitating ``kinks'' in 5D were investigated by Gremm and other workers \cite{Gremm,KT2,Giovannini2}. In a different construction, Kadosh et al \cite{KDP} obtained cosmologically plausible time dependent DW configurations by adopting a FLRW, maximally 3-symmetric ansatz for the 5D geometry, with crucial implications.

Here we have considered a thick brane \cite{thick1,thick2,GTV} in a 5D bulk, with a bulk scalar having self-interaction, a general exponential warp factor which depends both on time as well as on the extra-dimensional coordinate (as considered in \cite{GC1}) and a time-dependent extra dimension. To build a model
for a time-dependent process of localization of gravity, we have considered a time-dependent warp factor. During the early
phase of cosmological evolution (the high energy regime), the universe underwent a very rapid expansion, and therefore it may be presumed that the process of localization of gravity was time-dependent. This can be described by the time-dependent warp factor, which in turn gives rise to a time-varying bulk cosmological constant, so that the scale of the extra dimension is also time-dependent. The gravitational force law was sensitive to the background cosmological expansion \cite{DGP1,DGP2,Deffayet,DFGK1,DFGK2}, and the 4D Newton's constant was time-varying. The effective cosmological constant of the 4D hypersurface, which is determined by the curvature of the bulk metric, is time-varying due to the time-dependent process of localization, and this curvature controls the 4D Newton's constant. The dynamical evolution of the brane is determined by the Einstein equations for the combined brane-bulk system and the dynamics of the gravitational field is modified by the presence of the time-dependent warp factor. If this modification takes place at today's Hubble scale, $H_{0}$, then it is expected that the gravitational force law will get altered at distance scales much smaller than $H_{0}^{-1}$ \cite{Lue}, leading to an expansion history which can be reproduced by a dynamical dark energy. Somewhere around the end of inflation, the time-dependent process of localization of gravity came to an end, leading to the finite 4D Newton's constant in the present universe. This can be related to the infinite extent of the fifth dimension and is possible if the universe is inflating at early times. The warp factor is regular at the boundary of the 3-brane and is well-behaved at infinite distance from the brane. The brane tension is induced
both by the bulk  cosmological constant, as well as the curvature related to the expansion of the (3+1) spacetime and
the physical universe eventually accelerates. For an AdS bulk, the zero-mode graviton may not be localized on the
brane \cite{Neupane}.

The plan of our paper is as follows: After presenting the mathematical preliminaries in Section II, we have constructed the 5D field equations in Section III, and examined the status of the energy conditions in the bulk. We have derived the conditions for the bulk spacetime to be isotropic,
and have determined the nature of the bulk matter and the warping function. By inspecting the field equations for the isotropic bulk, we have assumed a definite correlation between the warping function and the extra-dimensional scale factor, following the results of supersymmetric brane models \cite{Gher1,Gher2}. In Section IV, we assumed a vanishing flux term for the generalized metric, signifying confinement of matter or energy (or both) within the 4D hypersurface, and have discussed the possibility of having a conformally flat bulk and the requirements for a
stabilized bulk. In Section V, we consider the flux term to be non-zero, which therefore represents the condition when
gravity becomes essentially higher dimensional. In this regime, for the most general warped metric in 5D, we have assumed
that exotic matter in the form of a scalar field with self-interaction potential (minimally coupled to gravity)
propagates in the bulk along with the gravitons. Finally, the summary and conclusions are presented in Section VI.

\section{Mathematical preliminaries}
Let us consider the generalized 5D action in presence of a real scalar field $\psi$ in the bulk which depends on time and also propagates in the bulk with potential $V(\psi)$, minimally
coupled to gravity:
\begin{equation}\label{01}
S= \int d^{5}x \sqrt{\bar{g}}\left[\frac{1}{2\bar{\kappa}^{2}}(\bar{R}-2\bar{\Lambda}) + \bar{L}_{B} +
\frac{1}{2} g^{AB}\nabla_{A} \psi \nabla_{B} \psi - V(\psi) \right] + \int d^{4}x \sqrt{-g}L_{m},
\end{equation}
supplemented with the brane curvature term
\begin{equation}\label{01a}
M_{P}^{2}\int d^{4}x \sqrt{-g}R,
\end{equation}
where $\bar{g}_{AB}$ is the 5D metric of signature (+ - - - -), $\bar{\Lambda}$ is the bulk cosmological constant and
$\bar{R}$ is the 5D scalar curvature. The Lagrangian density $L_{m}$ represents all other contribution to the action which are not strictly gravitational, including the contribution of the matter fields localized on the brane and any interaction
between the brane and the bulk, and $\bar{L}_{B}$ represents the contribution to the action arising purely due to bulk matter. The constant $\bar{\kappa}$ is related to the 5D Newton's constant $G_{(5)}$ and the 5D
reduced Planck mass $M_{(5)}$ by
\begin{equation}\label{02}
\bar{\kappa}^{2}= 8\pi G_{(5)}= M^{-3}_{(5)}.
\end{equation}
The 5D field equations are
\begin{equation}\label{04}
\bar{G}_{AB} = - \bar{\Lambda} \bar{g}_{AB} + \bar{\kappa}^{2} \bar{T}_{AB}
\end{equation}
where $\bar{G}_{AB}$ is the 5D Einstein tensor and $\bar{T}_{AB}$ represents the 5D energy-momentum tensor. The equation
of motion for the bulk scalar field $\psi$ is given by
\begin{equation}\label{04a}
\Box_{(5)} \psi + \frac{\partial V}{\partial \psi} = 0,
\end{equation}
where $\Box_{(5)}$ is the 5D covariant d'Alembertian operator. The energy-momentum tensor for a minimally coupled \textbf{real} scalar
field with potential is of the type
\begin{equation}\label{04b}
\bar{T}^{J(scalar)}_{I}= \partial_{I}\psi \partial^{J}\psi - \delta^{J}_{I} \left( \frac{1}{2}\bar{g}^{KL}\partial_{K}\psi \partial_{L}\psi
- V(\psi) \right).
\end{equation}

Warped spacetimes \cite{Carot} constitute a wide variety of exact solutions to the field equations and have been used to
solve the hierarchy problem and also recover Newtonian gravity in the low energy effective theory of gravity \cite{rs}.
We therefore choose a 5D metric ansatz given by
\begin{equation}\label{05}
dS^2 =e^{2f(t,y)}\left(dt^2 - R^2(t)(dr^2 + r^2d\theta^2 + r^2sin(\theta)^2d\phi^2)\right) - A^2(t,y)dy^2
\end{equation}
where $y$ is the coordinate of the fifth dimension, $t$ denotes the conformal time and the function $A(t,y)$ parametrizes
the scale of the extra dimension at different times and at different locations in the bulk. The "warping function", $f$,
is a smooth function, and $e^{2f(t,y)}$ is the time-dependent warp factor. The observed universe is represented by the
hypersurface at $y=y_{0}$.

\section{Field equations and energy conditions in the bulk}
The non-vanishing components of the 5D Einstein tensor for the space-time under consideration are
\begin{equation}\label{06a}
\bar{G}^{t}_{t}=\frac{3}{e^{2f}} \left( \frac{\dot{R}^2}{R^2} + \frac{2\dot{R}\dot{f}}{R} + \dot{f}^2
+ \frac{\dot{f}\dot{A}}{A} + \frac{\dot{R}}{R}\frac{\dot{A}}{A} \right) - \frac{3}{A^2}\left( 2f^{\prime 2}
+ f^{\prime\prime} -\frac{f^{\prime}A^{\prime}}{A} \right),
\end{equation}
\begin{equation}\label{06b}
\bar{G}^{t}_{y}=-\frac{3}{e^{2f}} \left( (\dot{f})^{\prime} - \frac{\dot{A}}{A}f^{\prime} \right),
\end{equation}
\begin{equation}\label{06c}
\bar{G}^{y}_{t}= \frac{3}{A^2} \left( (\dot{f})^{\prime} - \frac{\dot{A}}{A}f^{\prime} \right),
\end{equation}
\begin{equation}\label{06d}
\bar{G}^{y}_{y}= \frac{3}{e^{2f}} \left( \frac{\ddot{R}}{R} + \frac{\dot{R}^2}{R^2} + \frac{3\dot{R}\dot{f}}{R}
+ \dot{f}^2 + \ddot{f} \right)- \frac{6f^{\prime 2}}{A^2},
\end{equation}
and
\begin{equation}\label{06e}
\bar{G}^{I}_{I}=\frac{1}{e^{2f}} \left( \frac{2\ddot{R}}{R} + \frac{\dot{R}^2}{R^2} + \frac{4\dot{R}\dot{f}}{R}
+ \dot{f}^2 + 2\ddot{f} + \frac{\dot{f}\dot{A}}{A} + \frac{2\dot{R}}{R}\frac{\dot{A}}{A} + \frac{\ddot{A}}{A} \right)
- \frac{3}{A^2}\left( 2f^{\prime 2} + f^{\prime\prime} -\frac{f^{\prime}A^{\prime}}{A} \right).
\end{equation}
Above, a dot represents differentiation with respect to the conformal time $t$ and a prime stands for differentiation
with respect to the fifth coordinate $y$ and $I=r,\theta,\phi$. The components of the bulk stress-energy tensor are given by
\begin{center}
$\bar{T}^{t}_{t}=\bar{\rho},\qquad\qquad \bar{T}^{I}_{I}=-\bar{P},\qquad\qquad \bar{T}^{y}_{y}=-\bar{P}_{y},
\qquad\qquad \bar{T}^{t}_{y}=-\bar{Q}$.
\end{center}
where $\bar{\rho}$, $\bar{P}$ and $\bar{P}_{y}$ are the energy density and pressure in the bulk, with $\bar{Q}$ representing
a non-trivial flux term. The non-zero $\bar{G}^{t}_{y}$ term with the corresponding $\bar{T}^{t}_{y}$ component, indicates
a flux of matter or energy (or both) from the 4D hypersurface into the bulk along the extra dimension. The energy scale
necessary to ensure such an access is few hundred GeV. For such a generic bulk, the corresponding energy-momentum
tensor may not be diagonalizable, although for "physically reasonable" matter content on the hypersurface \cite{Wald}
(and may be in the bulk), we expect the stress-energy tensor to be diagonalizable. In our previous work with time-dependent warp factor \cite{GC1}, it was
found that the validity of the strong energy condition (SEC) in the bulk is governed by the nature of warping, as well as
by the effect of the extra dimension. Keeping in mind the possibility of a negative pressure component of matter-energy
(dark energy or cosmological constant), we require at least the weak energy condition (WEC) and the null energy
condition (NEC), to be obeyed. This implies that we must have
\begin{equation}\label{07}
\bar{\rho}\geq 0 \qquad\qquad \textrm{and} \qquad\qquad \bar{\rho}+\bar{P}_{i}\geq 0 \qquad\qquad (i=1,2,3,4).
\end{equation}
Assuming $8 \pi G_{(5)}=1$ in the remaining analysis, we obtain the field equations in 5D as
\begin{equation}\label{08a}
\bar{G}^{t}_{t}= - \bar{\Lambda} + \bar{\rho},
\end{equation}
\begin{equation}\label{08b}
\bar{G}^{I}_{I}= - \bar{\Lambda} - \bar{P},
\end{equation}
\begin{equation}\label{08c}
\bar{G}^{y}_{y}= - \bar{\Lambda} - \bar{P}_{y},
\end{equation}
and
\begin{equation}\label{08d}
\bar{G}^{t}_{y}= - \bar{\Lambda} - \bar{Q}.
\end{equation}
The complicated nature of the various components of the Einstein tensor makes it difficult for us to comment directly
on the status of the energy conditions from the above equations. To proceed further, let us write down the energy
conditions in terms of the energy eigen values (say, $a_{i}$), which must be real for matter-energy obeying the WEC
\cite{Hall1,Hall2}. These eigen values are given by the roots of the characteristic equation for the energy-momentum tensor:
\begin{equation}\label{09}
\textrm{det}(\bar{T} - a I) = 0.
\end{equation}
This implies that
\begin{center}
$\left|
  \begin{array}{ccccc}
    - \bar{\rho} + a & 0 & 0 & 0 & \bar{Q} \\
    0 & \bar{P}+ a & 0 & 0 & 0 \\
    0 & 0 & \bar{P}+ a & 0 & 0 \\
    0 & 0 & 0 & \bar{P}+ a & 0 \\
    \bar{Q} & 0 & 0 & 0 & \bar{P}_{y}+ a \\
  \end{array}
\right| = 0 $.
\end{center}
Solving this equation, we find that the eigen values are given by
\begin{eqnarray}\label{10}
  a_{0,4} &=& \frac{1}{2} \left[(\bar{\rho} - \bar{P}_{y})\pm\sqrt{ (\bar{\rho}+\bar{P}_{y})^2-4\bar{Q}^{2} } \right], \\
\textrm{and} \qquad \qquad \qquad  a_{1} &=& a_{2} = a_{3} = \bar{P}.
\end{eqnarray}
For these eigen values to be real, we must have,
\begin{equation}\label{11}
(\bar{\rho}+\bar{P}_{y})^2-4\bar{Q}^{2} \geq 0,
\end{equation}
along with the following restriction, so as to satisfy (\ref{07}), i.e.
\begin{equation}\label{12}
 a_{0} \geq 0 \qquad\qquad \textrm{and} \qquad\qquad a_{0} - a_{i} \geq 0  \qquad\qquad (i=1,2,3,4).
\end{equation}
The conservation of the energy-momentum tensor $\bar{T}^{a}_{b};a = 0$ leads us to the two equations
\begin{equation}\label{13}
\dot{\bar{\rho}} + 3 (\bar{\rho} + \bar{P}) \left( \dot{f} + \frac{\dot{R}}{R} \right) + (\bar{\rho}
+ \bar{P}_{y}) \frac{\dot{A}}{A} = 0,
\end{equation}
and
\begin{equation}\label{14}
\bar{P}^{\prime}_{y} + ( \bar{\rho} - 3\bar{P} + 4\bar{P}_{y} )f^{\prime} = 0.
\end{equation}
where $\frac{\dot{R}}{R}=H$, is the Hubble parameter on the brane.

\subsection{Conditions for an isotropic bulk}
For an isotropic bulk, we have $\bar{P}=\bar{P}_{y}$, so that condition (\ref{11}) reduces to
\begin{equation}\label{15}
(\bar{\rho}+\bar{P}) \geq |2\bar{Q}|,
\end{equation}
where $\bar{Q}$ may be positive or negative but $(\bar{\rho}+\bar{P})$ must be non-negative so as not to violate
the WEC. The condition $(\bar{\rho}+\bar{P}) > 0$, then corresponds to the \emph{high energy condition} for which
$|\bar{Q}| \neq 0$, so that gravity along with particles can access the higher dimensions. At low energies,
$|\bar{Q}| = 0$ and hence $(\bar{\rho}+\bar{P}) = 0$, which implies that in this condition, the energy density
of the bulk is sourced by a bulk cosmological constant.

At high energies, assuming positive flux for $\bar{Q}$, we can rewrite equation (\ref{13}) in the form
\begin{equation}\label{16}
\dot{\bar{\rho}} = - 2\bar{Q}\left( 3\dot{f} + \frac{3\dot{R}}{R} + \frac{\dot{A}}{A} \right),
\end{equation}
which represents the \emph{equation of continuity for this isotropic bulk}. For a surface enclosing a given
volume of the bulk (such that it encloses the 4D universe in the form of a 3-brane), since $\bar{Q}$ is positive,
the energy density inside this volume should decrease with time, if the quantity inside the brackets is positive.
Thus, the rate of decrease of energy density within the volume of the bulk spacetime enclosing the observed universe, depends on the rate of variation of the warping function, the rate of expansion of the observed universe and the normalized rate of variation of the extra-dimensional scale factor.

For the isotropic bulk, we get from equation (\ref{14}) the following result:
\begin{equation}\label{17}
(\bar{\rho} + \bar{P}) = - \frac{\bar{P}^{\prime}}{f^{\prime}}.
\end{equation}
The l.h.s. of the above equation will be non-negative under the following possibilities:

\begin{description}
  \item[(i)] $\bar{P}^{\prime}=0$, which implies that the pressure is a constant. This is possible when the bulk matter
  energy-momentum tensor is that of a perfect fluid representing dust. We can have $f^{\prime}$ to be either positive
  or negative, which means that the warp factor may be either growing or of decaying type. However, the decaying type
  is more favorable for cosmologically relevant solutions \cite{GC1}.
  \item[(ii)] $\bar{P}^{\prime}>0$, for which $f^{\prime}$ should necessarily be negative i.e. the warp factor is of
  decaying type, so that the WEC is valid.
  \item[(iii)] $\bar{P}^{\prime}<0$, in which case $f^{\prime}$ is positive and the warp factor is of growing type.
\end{description}
This is then the

Combining equations (\ref{08b}) and (\ref{08c}) for this isotropic bulk we obtain the relation
\begin{equation}\label{18}
\frac{1}{e^{2f}} \left( \frac{\ddot{R}}{R} + \frac{2\dot{R}^2}{R^2} + \frac{5\dot{R}\dot{f}}{R}
+ 2\dot{f}^2 + \ddot{f} - \frac{\dot{f}\dot{A}}{A} - \frac{2\dot{R}}{R}\frac{\dot{A}}{A}
- \frac{\ddot{A}}{A} \right) = - \frac{3}{A} \left( \frac{\partial}{\partial y} \left( \frac{f^{\prime}}{A} \right)\right).
\end{equation}
Since the l.h.s. involves only time-derivatives and the r.h.s. only the derivative with respect to the fifth coordinate,
without any loss of generality, we can assume that they are separately equal to some constant (say, $C$). From the r.h.s.
we can infer that the variation of the extra-dimensional scale factor along the direction of the extra dimensional coordinate at
different locations in the bulk, is related to the variation of the warp factor along the extra dimension. However,
accurate predictions can only be made if we know the exact functional relations.

Equating the constant $C$ to zero for the sake of simplicity, we obtain the following relation from the l.h.s. of (\ref{18})
\begin{equation}\label{18a}
\left( \frac{\ddot{R}}{R} + \frac{2\dot{R}^2}{R^2} + \frac{5\dot{R}\dot{f}}{R} + \dot{f}\left(2\dot{f}
- \frac{\dot{A}}{A} \right) + \ddot{f} - \frac{2\dot{R}}{R}\frac{\dot{A}}{A} - \frac{\ddot{A}}{A} \right) = 0.
\end{equation}

The \emph{RS scenario led to unnatural correlation between the bulk cosmological constant and the brane tensions}. However,
brane-bulk supersymmetry correctly correlates the brane tensions and the bulk cosmological constant in the \emph{supersymmetric RS models} \cite{Gher1,Gher2}. Under this condition, the metric (\ref{05}) will be a solution of the 5D Einstein equations, with the \emph{warp factor being related to the extra-dimensional scale factor}. For a stabilized bulk, $A(t,y)$ must settle down to a fixed value. A stabilized bulk with constant curvature and characterized by a negative cosmological constant is free from the appearance of nonconventional cosmologies \cite{Kanti}.

A \emph{spatially flat FRW-type braneworld} is appropriate for achieving cosmological solutions and can be embedded in any
constant curvature bulk \cite{maia2} e.g. a conformally flat bulk, in spite of representing a dynamical brane. For a
conformally flat bulk, the extra-dimensional scale factor can be a growing function of time \cite{KT2}. For such a bulk,
the resulting field equations are much simpler owing to the simplified geometry. However, \emph{a generalized bulk is usually not conformally flat}. In our model, we embed the spatially flat FRW braneworld in an arbitrary bulk.

Following the results of supersymmetric brane models, we assume the existence of brane-bulk supersymmetry \cite{Gher1,Gher2},
under which the warp factor is related to the extra-dimensional scale factor. Our assumption is justified because in
section IV, we do find that for low energy gravity (represented by the condition $\bar{G}^{t}_{y}=0=\bar{T}^{t}_{y}$),
there exists a definite correlation between the warp factor and the extra-dimensional scale factor. At this juncture
we want to point out that although our assumption is motivated by supersymmetry, no supersymmetry is involved in our
derivations.

Inspecting (\ref{18a}), we find that if we assume the following relation between the warp factor $f$ and
the extra-dimensional scale factor $A$, given by
\begin{equation}\label{18b}
2\dot{f} = \frac{\dot{A}}{A},
\end{equation}
then not only does it simplify equation (\ref{18a}), but in addition it helps us to derive a solution for the field
equations. From (\ref{18b}), the functional relation between $f$ and $A$ is found to be of the form

\begin{equation}\label{18d}
2f=\ln A - \ln C(y),
\end{equation}
which can be rewritten as follows
\begin{equation}\label{18e}
A(t,y) = C(y)e^{2f(t,y)},
\end{equation}
where $C(y)$ is and arbitrary function of $y$. Substituting (\ref{18b}) in (\ref{18a}) and simplifying, we obtain
\begin{equation}\label{18c}
\left( \frac{\ddot{R}}{R} + \frac{2\dot{R}^2}{R^2} + \frac{\dot{R}\dot{f}}{R} - 4\dot{f}^2 - \ddot{f}  \right) = 0.
\end{equation}
The above equation describes the nature of the time-variation of the warping function at a given location
in the bulk, specified by  $y=\textrm{constant}$. For a static brane, the condition $y=\textrm{constant}$ is sufficient to specify the
position of the brane along the extra dimension. However, for a dynamic brane, the deviation of the hypersurface from the
tangent plane will change with time \cite{GC1,maia2,maia3}, producing a time-varying ``bending effect''. Thus the position of the
hypersurface along the extra dimension changes with time. Geometrically, the extrinsic curvature of the hypersurface gives
us a measure of the deviation of the hypersurface from the tangent plane and the bending produces an observable effect in
the form of a smooth scalar function represented by the warp factor. Moreover, as this warp factor also depends on time,
it includes within itself the information regarding the time-dependent variation of the deviation of the brane. This variation in turn,
will affect the mechanism of localization of the fields on the brane.

Since the warp factor is smooth, let us assume that the deviation of the hypersurface changes slowly with time. During the
epoch when, $\ddot{f}=0$, equation (\ref{18c}) gets simplified further yielding the following solution \cite{Lake}:
\begin{eqnarray}
f(t,y) &=& \int \left. \frac{\dot{R}-\sqrt{33 \dot{R}^2 + 16 R \ddot{R}}}{8R}dt \right. + C_1(y) \label{19a}\\
       &=& C_2(t) + C_1(y), \label{19b}
\end{eqnarray}
where $C_1(y)$ is an arbitrary function of $y$ and $C_2(t)$ is determined by the specific form of $R(t)$. This means that, for the case $\ddot{f}=0$, the warp factor $e^{2f(t,y)}$ turns out to be of product type, which has already been considered by one of the authors in \cite{GC1}. However, for $\ddot{f}\neq 0$, we cannot derive a straightforward solution, without making an assumption on the functional form of the warp factor.

For the case $\ddot{R}=0$, if we have $\ddot{f}=0$, then the solution for the warping function is obtained as
\begin{eqnarray}
  f(t,y) &=& \frac{(1+\sqrt{33})}{8}\ln R + C_3(y),
\end{eqnarray}
where $C_3(y)$ is another arbitrary function of $y$. This is a very special case, but the analysis yields us a clear picture of the different conditions under which the warping function can be considered to be a separable function of $t$ and $y$.

\bigskip

We now proceed to investigate the nature of solution to the field equations, for the two different conditions: first at low
energies corresponding to the condition $\bar{G}^{t}_{y}=0=\bar{T}^{t}_{y}$, when gravity and particles are confined to
the four-dimensional hypersurface, and secondly at high energies, when gravity can access the bulk \cite{loclzn1,loclzn2,loclzn3,loclzn4,loclzn5,loclzn6,loclzn7,loclzn8,loclzn9,loclzn10,loclzn11}, the
condition being mathematically represented by $\bar{G}^{t}_{y}=\bar{T}^{t}_{y}\neq 0$.

\section{The condition $\bar{G}^{t}_{y}=0=\bar{T}^{t}_{y}$}
To prevent matter or energy flowing out of the brane along the fifth dimension, we require $\bar{T}^{t}_{y}=0$, which
implies that $\bar{G}^{t}_{y}=0$ and hence from (\ref{06b}) we obtain
\begin{equation}\label{20}
\dot{f}^{\prime}=\frac{\dot{A}}{A}f^{\prime}.
\end{equation}
Assuming that both $A$, $f$ and their first order derivatives are continuous, (\ref{20}) can be integrated to
give the result
\begin{equation}\label{21}
A(t,y)=\chi(y)f^{\prime}(t,y),
\end{equation}
indicating that the \emph{extra-dimensional scale factor is indeed related to the warp factor}, specifically, the extra-dimensional scale factor
at a given $y$ and $t$, depends on the way the warping function varies along the extra dimension at that instant at the
given location, and will be different at different locations in the bulk. The manner in which gravity is localized on the
4-dimensional hypersurface at different locations in the bulk (being monitored by the warp factor), will also be different
at different times. This is in agreement with the results of supersymmetric brane models as mentioned earlier, although no supersymmetry was involved in our derivation.

The metric (\ref{05}) now assumes the form:
\begin{equation}\label{21a}
dS^2 =e^{2f(t,y)}\left(dt^2 - R^2(t)(dr^2 + r^2d\theta^2 + r^2sin(\theta)^2d\phi^2)\right) - \chi^2(y)f^{\prime 2}(t,y)dy^2,
\end{equation}
Consequently, the Einstein equations for the above spacetime get reduced to the form
\begin{equation}\label{22}
\bar{\rho} - \Lambda_{(5)} =\frac{3}{e^{2f}} \left( \frac{\dot{R}}{R} + \dot{f} \right) \left( \frac{\dot{R}}{R}
+ \dot{f} + \frac{\dot{f}^{\prime}}{f^{\prime}} \right) - \frac{3}{\chi^{2}f^{\prime}}\left( 2f^{\prime}
- \frac{\chi^{\prime}}{\chi} \right),
\end{equation}
\begin{equation}\label{23}
\bar{P} + \Lambda_{(5)} = - \frac{1}{e^{2f}} \left( 2 \left( \frac{\ddot{R}}{R} + \frac{\dot{R}\dot{f}}{R} + \ddot{f} \right)
+ \left( \frac{\dot{R}}{R} + \dot{f} \right)^{2} + \frac{\dot{f}^{\prime}}{f^{\prime}} \left( \dot{f}
+ \frac{2\dot{R}}{R}\right) + \frac{\ddot{f}^{\prime}}{f^{\prime}} \right)
+ \frac{3}{\chi^{2}f^{\prime}}\left( 2f^{\prime} - \frac{\chi^{\prime}}{\chi} \right),
\end{equation}
and
\begin{equation}\label{24}
\bar{P}_{y} + \Lambda_{(5)} = - \frac{3}{e^{2f}} \left( \frac{\ddot{R}}{R} + \frac{\dot{R}^2}{R^2}
+ \frac{3\dot{R}\dot{f}}{R} + \dot{f}^2 + \ddot{f} \right) + \frac{6}{\chi^{2}}.
\end{equation}
Thus we have six unknowns ($f$, $R$, $\chi$, $\bar{\rho}$, $\bar{P}$ and $\bar{P}_{y}$) and three equations, leaving us
with the freedom of choosing three free functions. Choosing an isotropic bulk, we can reduce the number of unknowns to five. Subsequently, we can invoke the energy conditions and a specific equation of state to take care of the other two unknowns. That opens up innumerable number of possibilities, and we will report that analysis in a future paper \cite{SG3}. Instead, let us examine here whether the metric (\ref{21a}) can represent a conformally flat geometry or not.

\subsection{Nature of bulk geometry}
The non-zero components of the Weyl tensor for the line element (\ref{21a}), are given by
\begin{equation}\label{25}
C_{AB}^{AB}= \textrm{factor} \times \left(\frac{1}{e^{2f}}\left(\frac{\ddot{R}}{R} - \frac{\dot{R}}{R}\left(\dot{f}
+ \frac{\dot{R}}{R}\right) + \ddot{f} - \dot{f}^2 + \frac{\dot{f}^{\prime}}{f^{\prime}}\left(\frac{\dot{R}}{R}
+ 2\dot{f}\right) - \frac{\ddot{f}^{\prime}}{f^{\prime}}\right) \right)
\end{equation}
where, only the ``factor'' varies for the different components of the Weyl tensor. The condition for a conformally flat
bulk is known to be given by
\begin{equation}\label{25a}
C_{AB}^{AB}=0.
\end{equation}
If (\ref{25a}) is to be satisfied, then (\ref{25}) implies that we must have
\begin{equation}\label{25b}
\frac{1}{e^{2f}}\left(\frac{\ddot{R}}{R} - \frac{\dot{R}}{R}\left(\dot{f} + \frac{\dot{R}}{R}\right) + \ddot{f} - \dot{f}^2
+ \frac{\dot{f}^{\prime}}{f^{\prime}}\left(\frac{\dot{R}}{R} + 2\dot{f}\right)
- \frac{\ddot{f}^{\prime}}{f^{\prime}}\right)=0. \nonumber
\end{equation}
In that case $f(t,y)$ must be of the form $f(t,y)=F_{1}(t)\,\times\, \textrm{constant}$, since $f(t,y)$ also depends on $y$.
In fact solving (\ref{25b}) for $f(t,y)$ \cite{Lake}, we find that
\begin{equation}\label{26}
f(t,y)=F_{1}(t)F_{2}(y) \qquad \textrm{where} \qquad F_{2}(y)= \textrm{constant},
\end{equation}
$F_{1}(t)$ being an arbitrary function of time. But (\ref{26}) basically implies an unwarped metric. In that case the
correlation (\ref{21}) also becomes invalid since the extra-dimensional scale factor becomes trivial and the concept of the fifth dimension loses sense. Therefore, a geometry represented by a line element of the type (\ref{21a}) with a generalised warp factor and a generalized scale factor
for the extra dimension, \emph{will not be conformally flat}, even if $\bar{G}^{t}_{y}=0$.

We are tempted to leave the following comment at this point: \emph{Only with a separable warp factor}, as for example done by \cite{GC1,ahmedgrad1,ahmedgrad2}, we will have a conformally flat geometry for a 5D line element with a time-dependent warp factor.

\subsection{Requirements for a stabilized bulk}
For the bulk to be stabilised, we need $\dot{A}$ = 0, which implies that $A = A(y)$. In that case it follows that for
$\bar{T}^{t}_{y} = 0$, we need $\dot{f}^{\prime} = 0$, which will be satisfied provided either $f = f(t)$ or
$f = f(y)$. But $f=f(t)$ corresponds to the unwarped case for which we can simply absorb $f(t)$ inside the 4D metric.
For $f=f(y)$, we recover the usual RS scenario. When $f''=0 \Rightarrow f'=\textrm{constant}$, we have $f= const \times y$, corresponding to a de Sitter
3-brane embedded in a AdS bulk. If the bulk is of constant curvature, then the 5D Weyl tensor vanishes and the bulk is
characterized by a negative cosmological constant with $\rho_{B} = -P_{B}$.

If however, the warp factor is of product form, as discussed in an earlier paper \cite{GC1}, an assumption that transforms
the Einstein equations into a simpler form, which can be solved by straightforward calculations, then the energy-momentum
tensor has only the $\bar{T}^{y}_{y}$ component other than the components on the brane. This set-up also corresponded to a
stabilized bulk. The validity of SEC in the bulk for such a metric is governed by the nature of warping and the effect of
the extra dimension. It was shown in \cite{GC1} that the WEC is valid for an isotropic bulk for a certain form of the 4D scale
factor and the time-dependent warp factor The 4D universe is initially decelerated, but makes a transition to an
accelerated phase at later times, leading to an interpretation of dynamical dark energy which is related to the
time-dependent warp factor.

\section{Gravity at high energies when $\bar{G}^{t}_{y}=\bar{T}^{t}_{y}\neq 0$}

We now consider that the energy scale is in the GeV range, so that particles can escape into the extra dimension. Here we incorporate the correlation between the warping function and the extra-dimensional scale factor that we derived in Section III following the results of supersymmetric brane models, from equation (\ref{18e}), into our metric ansatz in (\ref{05}). In that case, the 5D metric assumes the form
\begin{equation}\label{05new}
dS^2 =e^{2f(t,y)}\left(dt^2 - R^2(t)(dr^2 + r^2d\theta^2 + r^2sin(\theta)^2d\phi^2)\right) - e^{2f(t,y)}dy^2,
\end{equation}
where we have redefined the $y$-coordinate by replacing $dy \longrightarrow (C(y))^{-1/2}dy$, the function $C(y)$ being the same as the one which appeared in equation (\ref{18e}).

Further, we assume that the 5D bulk is characterised by a scalar field $\psi(t,y)$ with self-interaction potential $V(\psi)$, minimally
coupled to gravity \cite{RPU,Giovannini}. With the bulk metric now given by (\ref{05new}), the scalar field equation of motion reduces to the following form:
\begin{equation}\label{B3}
\ddot{\psi} - \psi'' + \dot{\psi}(3Q + 3H) - 3P\psi' + \Gamma^2 \frac{\partial V}{\partial \psi} = 0,
\end{equation}
where $H$ ($= \frac{\dot{R}}{R}$) is the Hubble parameter on the brane, and we have introduced the following redefinition of parameters

\begin{description}
  \item[(i)] $e^{2f(t,y)} = \Gamma^2 \Rightarrow f(t,y) = \ln \Gamma$,
  \item[(ii)] $P(t,y) = \frac{\partial f}{\partial y} = \frac{\partial \ln \Gamma}{\partial y}$,
  \item[(iii)] $Q(t,y) = \frac{\partial f}{\partial t} = \frac{\partial \ln \Gamma}{\partial t}$.
\end{description}

Consequently the Einstein equations in the bulk appear as follows:
\begin{eqnarray}
6 (H^2 + 3HQ + 2Q^2 - P^2 - P') =  \dot{\psi}^2 + \psi'^2 + 2\Gamma^2 \left( V - \bar{\Lambda} \right), \label{B4a}\\
6QP - 3(Q' + \dot{P}) = 2\dot{\psi} \psi^{\prime},  \qquad\qquad\qquad\qquad\qquad \label{B4b}\\
6( \dot{H} + 2H^2 + 3HQ + \dot{Q} + Q^2 - 2P^2 ) = \dot{\psi}^2 - 3\psi'^{2} - 2 \Gamma^2 \left( V - \bar{\Lambda} \right), \qquad \label{B4c}\\
4\dot{H} + 6( H^2 + 2HQ + Q^2 + \dot{Q} - P^2 - P' ) = \dot{\psi}^2 - \psi'^{2} - 2 \Gamma^2 \left( V - \bar{\Lambda} \right). \qquad \label{B4d}
\end{eqnarray}

Subtracting (\ref{B4d}) from (\ref{B4c}), we get
\begin{equation}\label{B5}
\psi'^{2} = 3P^2 - 3P' - \dot{H} - 3H(H + Q).
\end{equation}
On the other hand, adding (\ref{B4a}) and (\ref{B4d}), we obtain
\begin{equation}\label{B6}
\dot{\psi}^2 = 2\dot{H} + 3H(2H + 5Q) + 9Q^2 + 3\dot{Q}.
\end{equation}
Equations (\ref{B5}) and (\ref{B6}) indicate that the scalar field $\psi$ is independent of $\bar{\Lambda}$.

Let us now consider the following form of the warp factor:
\begin{equation}\label{B7}
e^{2f(t,y)} = \Gamma^2 = \frac{1}{1 + \alpha^2 (t + y)^2},
\end{equation}
where $\alpha$ is a dimensionless parameter. In that case, the warping function $f(t,y)$ and the variables $P(t,y)$ and $Q(t,y)$ are given by
\begin{eqnarray}
  f(t,y) = - \frac{1}{2} \ln[1 + \alpha^2 (t + y)^2], \label{B8_1} \\
  Q(t,y) = P(t,y) = - \frac{\alpha^2 (t + y)}{1 + \alpha^2 (t + y)^2}. \label{B8_2}
\end{eqnarray}
Further, we assume that the scalar field has the form
\begin{equation}\label{B9}
\psi(t,y) = \sqrt{3}\arctan[\alpha(t + y)],
\end{equation}
such that the warp factor and the scalar field are related by
\begin{equation}\label{B(_1}
e^{2f(t,y)} = \Gamma^2 = \frac{1}{\sqrt{3 \alpha}} \frac{\partial \psi}{\partial u},
\end{equation}
where $u=(t+y)$. Basically, this is similar to the time shifted solution derived by Giovannini in \cite{Giovannini} by generalizing the static maximally 4-symmetric solution in \cite{Giovannini2} to the time-dependent case by the simple coordinate redefinition $y \rightarrow y + t$. This choice satisfies the field equations, with the bulk cosmological constant $\bar{\Lambda}$ being determined in terms of $H$, the Hubble parameter of the brane at the particular position in the bulk. To determine the scalar field potential $V(\psi)$, we plug in (\ref{B7}), (\ref{B8_1}), (\ref{B8_2}) and (\ref{B9}) into the equation of motion (\ref{B3}), which then yields
\begin{equation}\label{B10}
V(\psi) = -3 \sqrt{3} \alpha \int H d\psi.
\end{equation}
For the simple case, $H=\textrm{constant}$, i.e. $H$ is independent of time, (\ref{B10}) can be integrated to yield
\begin{equation}\label{B11}
V(\psi) = -3 \sqrt{3} \alpha H \psi.
\end{equation}
Thus, whether $H$ is a constant or not, the potential of the scalar field source of gravity at a given time and a given position along the fifth dimension, is related to the Hubble parameter on the brane at that location in the bulk, i.e., it is related to the rate of expansion of the brane universe at that position in the bulk.

We note here that the line element (\ref{05new}), is similar to the line element (2.1) in the paper \cite{Giovannini}, by Giovannini, which corresponds to time-dependent gravitating solitons in 5D warped spacetimes. The parameter $\epsilon$ appearing in the correlation for the warp factors in equation (2.11) of \cite{Giovannini}, is equal to unity in our case, since the two warp factors are identical in our line element (\ref{05new}). In his paper, Giovannini considered a Minkowski brane, but here we have considered a spatially flat FRW brane. The scalar field potential $V(\psi)$ in our case is non-zero, as is found in equation (\ref{B10}), and is related to the Hubble parameter on the brane at the particular value of extra-dimensional coordinate $y$. On the other hand, with a Minkowski brane in \cite{Giovannini}, the condition $\epsilon\rightarrow1$ reduced the solution to an essentially time-dependent one with vanishing potential (i.e. $V(\psi)=0$). The solution for $\epsilon=1$ in \cite{Giovannini} had no static analog, since the time-dependence was essential. In our case too, the time-dependence is essential.

\section{Summary and conclusions}
In this paper, we have considered five-dimensional warp product space-times with a time-dependent warp factor and a
non-compact fifth dimension. The warp factor reflects the confining role of the bulk cosmological constant to localize
gravity at the brane through the curvature of the bulk. Since this process of localization is expected to have some
time-dependence during the early phase of cosmological evolution, we have considered a generalized warp factor which
depends both on time as well as on the extra-dimensional coordinate. The extra-dimensional scale
factor is also a function of time and of the extra coordinate, so that we are dealing with a very general type of bulk.
The eigen values for the energy-momentum tensor are evaluated and the condition for the WEC and NEC to hold in the bulk
are discussed. Combining these conditions with the conservation equations, we have been able to comment on the nature
of matter-energy tensor for this bulk and the type of warping function.

At low energies, when gravity and particles remain confined to the brane, corresponding to the case $\bar{T}^t_y=0$, we
found that the extra-dimensional scale factor depends on the variation of the warping function along the extra dimension.
In general, the bulk is not conformally flat, although it can be so if the extra-dimensional scale factor is only a
function of time or the warp factor is of product type, being separable into a time-dependent and a extra dimension dependent
part. When the variation of the warping function along the extra dimension assumes a constant value, i.e., $f'=\textrm{constant}$, we have a de Sitter 3-brane embedded in a AdS bulk. For a stabilized bulk of constant curvature, our model reduces to the standard RS scenario, with the bulk characterized by a negative
cosmological constant.

Finally, at high energies when $\bar{T}^t_y\neq 0$, we have considered the generalized warped metric for the bulk to be sourced by a bulk scalar field. Considering a self-interacting scalar field, and with a given choice for the warping function, we have determined the solution to the scalar field, and the scalar field potential from the field equations. We find that both the bulk cosmological constant and the potential of the scalar field source of gravity at a given time and a given position along the fifth dimension, is related to the Hubble parameter on the brane at that location in the bulk, i.e., it is related to the rate of expansion of the brane universe at that position in the bulk.

\section*{Acknowledgments}
The major part of this work was done in IUCAA, India under the associateship programme. SG gratefully acknowledges
the warm hospitality and the facilities of work at IUCAA. Our thanks to Professor Asit Banerjee for his comments and
suggestions at the Relativity and Cosmology Research Center, Jadavpur University and to Professor Subenoy Chakraborty
for the useful discussions. We are also thankful to the anonymous referee for the constructive suggestions.


\begin{thebibliography}{}
\bibitem{AKA1} K. Akama, Lect. Notes Phys. \textbf{176}, 267 (1982) [arXiv:hep-th/0001113]
\bibitem{Rubakov} V. A. Rubakov and M. R. Shaposhnikov, Phys. Lett. B \textbf{125}, 136 (1983)
\bibitem{GSW} M. Green, J. Schwarz and E. Witten, \emph{Superstring Theory}, Vol. I and II (Cambridge University Press,
    Cambridge, 1987)
\bibitem{Pol} J. Polchinski, String Theory, Vol. I and II, (Cambridge University Press, Cambridge, 1998)
\bibitem{Lukas1} A. Lukas, B. A. Ovrut, K. S. Stelle, and D. Waldram, Phys. Rev. D \textbf{59}, 086001 (1999)
\bibitem{Lukas2} A. Lukas, B. A. Ovrut, K. S. Stelle, and D. Waldram, Nucl. Phys. B \textbf{552}, 246 (1999)

\bibitem{ADD1} N. Arkani-Hamed, S. Dimopoulos and G. Dvali, Phys. Lett. B \textbf{429}, 263 (1998)
\bibitem{ADD2} N. Arkani-Hamed, S. Dimopoulos and G. Dvali, Phys. Rev. D \textbf{59}, 086004 (1999)
\bibitem{ADD3} G. Dvali and M. Shifman, Phys. Lett. B \textbf{396}, 64 (1997)
\bibitem{ADD4} I. Antoniadis, N. Arkani-Hamed, S. Dimopoulos and G. Dvali, Phys. Lett. B \textbf{436}, 257 (1998)
\bibitem{rs} L. Randall and R. Sundrum, Phys. Rev. Lett. \textbf{83}, 3370 (1999) ; \emph{ibid.} 4690

\bibitem{lrr} R. Maartens and K. Koyama, \emph{Living Rev. Relativity} \textbf{13}, 5 (2010); arXiv:1004.3962[hep-th]
\bibitem{MS} \emph{Brane World: New Perspectives in Cosmology}, Ed. K. Maeda and M. Sasaki [Prog. Theor. Phys. Suppl.
    \textbf{148}, 181 (2002)]
\bibitem{Mannheim} P. D. Mannheim, \emph{Brane-Localized Gravity} (Singapore: World Scientific, 2005)
\bibitem{loclzn1} B. Bajc and G. Gabadadze, Phys. Lett. B \textbf{474}, 282  (2000)
\bibitem{loclzn2} J. Garriga and T. Tanaka, Phys. Rev. Lett. \textbf{84}, 2778 (2000)
\bibitem{loclzn3} S.B. Giddings, E. Katz and L. Randall, J. High Energy Phys. \textbf{0003}, 023 (2000)
\bibitem{loclzn4} E. Kh. Akhmedov, Phys. Lett. B \textbf{521}, 79 (2001)
\bibitem{loclzn5} I. Brevik, K. Ghoroku, S.D. Odintsov and M. Yahiro, Phys. Rev. D \textbf{66}, 064016 (2002)
\bibitem{loclzn6} M. Gogberashvili and P. Midodashvili, Europhys. Lett. \textbf{61}, 308 (2003)
\bibitem{loclzn7} M. Gogberashvili and D. Singleton, Phys. Rev. D \textbf{69}, 026004 (2004)
\bibitem{loclzn8} G. Gibbons, K. Maeda and Y. Takamizu, Phys. Lett. B \textbf{647}, 1 (2007)
\bibitem{loclzn9} N. Barbosa-Cendejas, A. Herrera-Aguilar, M. A. Reyes Santos and C. Schubert, Phys. Rev. D \textbf{77}, 126013 (2008)
\bibitem{loclzn10} R. Guerrero, A. Melfo, N. Pantoja and R. O. Rodriguez, Phys. Rev. D \textbf{81}, 086004 (2010)
\bibitem{loclzn11} L. J. S. Sousa, W. T. Cruz and C. A. S. Almeida, Phys. Lett. B \textbf{711}, 97 (2012)
\bibitem{RPU} C. Ringeval, P. Peter and J.-P. Uzan, Phys. Rev. D \textbf{65}, 044016 (2002)
\bibitem{bdl1} P. Binetruy, C. Deffayet and D. Langlois,  Nucl. Phys. B \textbf{565}, 269 (2000)
\bibitem{bdl2} P. Binetruy, C. Deffayet, U. Ellwanger and D. Langlois, Phys. Lett. B \textbf{477}, 285 (2000)
\bibitem{others1a} J. M. Cline, C. Grojean, and G. Servant, Phys. Rev. Lett. \textbf{83}, 4245 (1999)
\bibitem{others1b} J. M. Cline and J. Vinet, J. High Energy Phys. \textbf{0202}, 042 (2002)
\bibitem{others2a} P. J. Steinhardt, Phys. Lett. B \textbf{462}, 41 (1999)
\bibitem{others2b} C. Csaki, M. Graesser, L. Randall, and J. Terning, Phys. Rev. D \textbf{62}, 045015 (2000)
\bibitem{others3a} E. E. Flanagan, S. H. H. Tye, and I. Wasserman, Phys. Rev. D \textbf{62}, 044039 (2000)
\bibitem{others3b} H. Stoica, S. H. H. Tye, and I. Wasserman, Phys. Lett. B \textbf{482}, 205 (2000)
\bibitem{others4a} R. Maartens, D. Wands, B. A. Bassett, and I. Heard, Phys. Rev. D \textbf{62}, 041301 (2000)
\bibitem{others4b} E. J. Copeland, A. R. Liddle, and J. E. Lidsey, Phys. Rev. D \textbf{64}, 023509 (2001)
\bibitem{others4c} P. R. Ashcroft, C. van de Bruck, and A. C. Davis, Phys. Rev. D \textbf{66}, 121302 (2002)
\bibitem{AKA2} K. Akama and T. Hattori, Class. Quantum Grav. \textbf{30}, 205002 (2013)
\bibitem{varmod1} P. Brax, C. van de Bruck, A. C. Davis and C. S. Rhodes, Phys. Rev. D, \textbf{67}, 023512 (2003)
\bibitem{varmod2} T. Damour and K. Nordtvedt, Phys. Rev. Lett. \textbf{70}, 2217 (1993)
\bibitem{varmod3} D. I. Santiago, D. Kalligas, and R. Wagoner, Phys. Rev. D \textbf{58}, 124005 (1998)
\bibitem{early1} T. Barreiro, B. de Carlos, and E. J. Copeland, Phys. Rev. D \textbf{58}, 083513 (1998)
\bibitem{early2} K. Choi, H.B. Kim, and H.D. Kim, Mod. Phys. Lett. A \textbf{14}, 125 (1999)
\bibitem{early3} G. Huey, P.J. Steinhardt, B.A. Ovrut, and D. Waldram, Phys. Lett. B \textbf{476}, 379 (2000)
\bibitem{GW} W. D. Goldberger and M. B. Wise, Phys. Rev. Lett. \textbf{83}, 4922 (1999); Phys. Rev. D \textbf{60}, 107505 (1999)

\bibitem{GPT} J. Garriga, O. Pujolas and T. Tanaka, Nucl. Phys. B \textbf{605}, 192 (2001)
\bibitem{powerlaw} J. Garriga, O. Pujolas and T. Tanaka, Nucl. Phys. B \textbf{655}, 127 (2003)
\bibitem{KT1} K. Koyama and K. Takahashi, Phys. Rev. D \textbf{68}, 103512 (2003)
\bibitem{KKOP} P. Kanti, I. I. Kogan, K. A. Olive, and M. Pospelov, Phys. Rev. D \textbf{61}, 106004 (2000)

\bibitem{scalar1} T. Tanaka and X. Montes, Nucl. Phys. B \textbf{582}, 259 (2000)
\bibitem{scalar2} P. Kanti, S. Lee and K. A. Olive, Phys. Rev. D \textbf{67}, 024037 (2003)
\bibitem{scalar3} R. Koley and S. Kar, Class. Quantum Grav. \textbf{22}, 753 (2005)
\bibitem{scalar4} R. Koley and S. Kar, Class. Quantum Grav. \textbf{24}, 79 (2007)
\bibitem{scalar5} R. Koley and S. Kar, Phys. Lett. B \textbf{623}, 244 (2005)
\bibitem{scalar6} S. Pal and S. Kar, Gen. Relativ. Gravit. \textbf{41}, 1165 (2009)

\bibitem{MK} S. Mukohyama and L. Kofman, Phys. Rev. D \textbf{65}, 124025 (2002)
\bibitem{DGP1} G. Dvali, G. Gabadadze and M. Porrati, Phys. Lett. B \textbf{484}, 112 (2000)
\bibitem{DGP2} G. Dvali, G. Gabadadze and M. Porrati, Phys. Lett. B \textbf{485}, 208 (2000)
\bibitem{Deffayet} C. Deffayet, Phys. Lett. B \textbf{502}, 199 (2001)
\bibitem{rs2cosmo1a} R. Gregory, V. A. Rubakov, S. M. Sibiryakov, Phys. Rev. Lett. \textbf{84}, 5928 (2000)
\bibitem{rs2cosmo1b} P. Kraus, J. High Eneryg Phys. \textbf{12}, 011 (1999)
\bibitem{rs2cosmo1c} D. Ida, J. High Energy Phys. \textbf{0009}, 014 (2000)
\bibitem{rs2cosmo2a} Y. Shtanov, hep-th/0005193
\bibitem{rs2cosmo2b} J. Garriga and M. Sasaki, Phys. Rev. D \textbf{62}, 043523 (2000)
\bibitem{rs2cosmo2c} H. A. Chamblin and H. S. Reall, Nucl. Phys. B \textbf{562}, 133 (1999)
\bibitem{rs2cosmo2d} N. Kaloper, Phys. Rev. D \textbf{60}, 123506 (1999)
\bibitem{rs2cosmo2e} S. Nojiri and D. Odintsov, Phys. Lett. B \textbf{484}, 119 (2000)
\bibitem{BCG} F. Bonjour, C. Charmousis and R. Gregory, Class. Quantum Grav. \textbf{16}, 2427 (1999)

\bibitem{Gremm} M. Gremm, Phys. Lett. B \textbf{478}, 434 (2000)
\bibitem{Giovannini2} M. Giovannini, Phys. Rev. D \textbf{64}, 064023 (2001)
\bibitem{Bronnikov1} K. A. Bronnikov and B.E. Meierovich, Grav. \& Cosmol. \textbf{9}, 313 (2003)
\bibitem{Bronnikov2} K. A. Bronnikov, S. B. Fadeev and A. V. Michtchenko, Gen. Relativ. Gravit. \textbf{36}, 1527 (2004)
\bibitem{DFGK1} O. DeWolfe, D.Z. Freedman, S.S. Gubser and A. Karch, Phys. Rev. D \textbf{62}, 046008 (2000)
\bibitem{DFGK2} D. Bazeia, A.R. Gomes and L. Losano, Int. J. Mod. Phys. A \textbf{24}, 1135 (2009)
\bibitem{ahmedgrad1}  A. Ahmed, B. Grzadkowski and J. Wudka, J. High Energy Phys. \textbf{1404}, 061 (2014)
\bibitem{ahmedgrad2}  A. Ahmed and B. Grzadkowski, J. High Energy Phys. \textbf{1301}, 177 (2013)
\bibitem{ahmedgrad3}  A. Ahmed, L. Dulny and B. Grzadkowski, Eur. Phys. J. C \textbf{74}, 2862 (2014)

\bibitem{Farakos1} K. Farakos and P. Pasipoularides, Phys. Rev. D \textbf{73}, 084012 (2006)
\bibitem{Farakos2} K. Farakos and P. Pasipoularides, Phys. Rev. D \textbf{75}, 024018 (2007)

\bibitem{SMS1} T. Shiromizu, K-I. Maeda and M. Sasaki, Phys. Rev. D \textbf{62}, 024012 (2000)
\bibitem{SMS2} R. Maartens, Phys. Rev. D \textbf{62}, 084023 (2000)
\bibitem{Greg1} P. Bowcock, C. Charmousis and R. Gregory, Class. Quantum Grav. \textbf{17}, 4745 (2000)
\bibitem{Greg2} R. Gregory and J. A. Harvey, Class. Quantum Grav. \textbf{20}, L231 (2003)
\bibitem{Greg3} G. Niz, A. Padilla and H. K. Kunduri, J. Cosmol. Astropart. Phys. \textbf{0804}, 012 (2008)
\bibitem{nogo1} P. K. Townsend and M. N. R. Wohlfarth, Phys. Rev. Lett. \textbf{91}, 061302 (2003)
\bibitem{nogo2} N. Ohta, Phys. Rev. Lett. \textbf{91}, 061303 (2003)
\bibitem{GC1} S. Guha and S. Chakraborty, Int. J. Theor. Phys. \textbf{51}, 55 (2012); Gen. Relativ. Gravit. \textbf{42}, 1739 (2010)
\bibitem{GK1a} S. Ghosh and S. Kar, Phys. Rev. D \textbf{80}, 064024 (2009)
\bibitem{GK1b} S. Ghosh, A. Dasgupta and S. Kar, Phys. Rev. D \textbf{83}, 084001 (2011)
\bibitem{Gog} M. Gogberashvili, A. Herrera-Aguilar and D. Malag´on-Morej´on, Class. Quantum Grav. \textbf{29}, 025007 (2012)
\bibitem{sugra1} G. W. Gibbons, H. Lu and C. N. Pope, Phys. Rev. Lett. \textbf{94}, 131602 (2005)
\bibitem{sugra2} W. Chen, Z. W. Chong, G. W. Gibbons, H. Lu and C. N. Pope, Nucl. Phys. B \textbf{732}, 118 (2006)
\bibitem{sugra3} K. Maeda, N. Ohta, M. Tanabe and R. Wakebe, J. High Energy Phys. \textbf{0906}, 036 (2009)
\bibitem{sugra4} P. Binetruy, M. Sasaki and K. Uzawa, Phys. Rev. D \textbf{80}, 026001 (2009) 
\bibitem{sugra5} G. W. Gibbons and K. Maeda, Phys. Rev. Lett. \textbf{104}, 131101 (2010)
\bibitem{sugra6} M. Minamitsuji and K. Uzawa, Phys. Rev. D \textbf{83}, 086002 (2011)
\bibitem{sugra7} K. Maeda and K. Uzawa, Phys. Rev. D \textbf{85}, 086004 (2012)

\bibitem{Bernardini} A. E. Bernardini, R. T. Cavalcanti and R. da Rocha, Gen. Relativ. Gravit. \textbf{47}, 1840 (2015)
\bibitem{BB} A. E. Bernardini, and O. Bertolami, Phys. Lett. B \textbf{726}, 512 (2013) 
\bibitem{BR}A. E. Bernardini and R. da Rocha, Adv. High Energy Phys. \textbf{2016}, 3650632 (2016)

\bibitem{Giovannini} M. Giovannini, Phys. Rev. D \textbf{76}, 124017 (2007)
\bibitem{KT2} A. Kehagias and K. Tamvakis, Phys. Lett. B \textbf{504}, 38 (2001)


\bibitem{KDP} A. Kadosh, A. Davidson and Elisabetta Pallante, Phys. Rev. D \textbf{86}, 124015 (2012)
\bibitem{thick1} A. K. Raychaudhuri and G. Mukherjee, Phys. Rev. Lett. \textbf{59}, 1504 (1987) 
\bibitem{thick2} V. Dzhunushaliev, V. Folomeev and M. Minamitsuji, Rep. Prog. Phys. \textbf{73}, 066901 (2010)
\bibitem{GTV} D. P. George, M. Trodden and R. R. Volkas, J. High Energy Phys. \textbf{0902}, 035 (2009)

\bibitem{Lue}  A. Lue, R. Scoccimarro and G. D. Starkmann, Phys. Rev. D \textbf{69}, 124015 (2004)
\bibitem{Neupane} I. P. Neupane, Phys. Rev. D \textbf{83}, 086004 (2011); Class. Quantum Grav. \textbf{26}, 195008 (2009)
\bibitem{Gher1} T. Gherghetta and A. Pomarol, Nucl.Phys. B \textbf{586}, 141 (2000)
\bibitem{Gher2} R. Altendorfer, J. Bagger and D. Nemeschansky, Phys. Rev. D \textbf{63}, 125025 (2001)
\bibitem{Carot} J. Carot and J. da Costa, Class. Quantum Grav. \textbf{10}, 461 (1993)
\bibitem{Wald} R. M. Wald, \emph{General Relativity} (University of Chicago Press, Chicago, 1984)
\bibitem{Hall1} G. S. Hall, J. Phys. A, \textbf{9}, 541 (1976)
\bibitem{Hall2} C. A. Kolassis, N. O. Santos and D. Tsoubelis, Class. Quantum Grav. \textbf{5}, 1329 (1988)

\bibitem{Kanti} P. Kanti, I. Kogan, K. A. Olive and M. Pospelov, Phys. Lett. B \textbf{468}, 31 (1999)
\bibitem{maia2} M. D. Maia, E. M. Monte, J. M. F. Maia and J. S. Alcaniz, Class. Quantum Grav. \textbf{22}, 1623 (2005)
\bibitem{maia3} M. D. Maia, E. M. Monte and J. M. F. Maia, Phys. Lett. B \textbf{585}, 11 (2004)
\bibitem{Lake} These calculations were done with the help of Maple program and GRTensor package: K. Lake and P. J. Musgrave,
    GRTensor (Queen's University, Kingston, 2003)
\bibitem{SG3} S. Guha and S. Das (in preparation).


\end{thebibliography}
\end{document}